\documentstyle [12pt]{article}
\makeatletter \oddsidemargin  0in \evensidemargin 0in
\textwidth16cm \RequirePackage[dvips]{graphicx} \textheight 20.5cm
\newcommand{\singlespacing}{\let\CS=\@currsize\renewcommand{\baselinestretch}{1.5}\tiny\CS}
\makeatletter \oddsidemargin  -.1in \evensidemargin -.1in
\newcommand{\doublespacing}{\let\CS=\@currsize\renewcommand{\baselinestretch}{1.35}\tiny\CS}

\def\@citex[#1]#2{\if@filesw\immediate\write\@auxout{\string\citation{#2}}\fi
  \def\@citea{}\@cite{\@for\@citeb:=#2\do
    {\@citea\def\@citea{,\linebreak[0]\hskip0pt plus .2em}%
      \@ifundefined{b@\@citeb}%
    {{\bf ?}\@warning{Citation `\@citeb' on page \thepage\space undefined}}%
      \hbox{\csname b@\@citeb\endcsname}}}{#1}}

\newtheorem{rule-def}[theorem]{Rule}
\begin{document}
\title{\bf Deletion, Bell's Inequality, Teleportation}
\author{Indranil Chakrabarty $^{1,3}$\thanks{Corresponding author:
E-Mail-indranilc@indiainfo.com }, \\ Nirman Ganguly $^{1,2}$,Binayak S. Choudhury$^{2}$ \\
$^1$Heritage Institute of Technology, Kolkata - 107, WestBengal, India.\\
$^2$Bengal Engineering and Science University, Shibpur, West
Bengal, India.\\
$^3$ Institute of Physics, Sachivalaya Marg, 751005 Bhubaneswar, Orissa, India }
\date{}
\maketitle{}
\begin{abstract}
In this letter we analyze the efficacy of the entangled output of
Pati-Braunstein  deletion machine [3] as a teleportation channel.
We analyze the possibility of it violating the Bell's inequality.
Interestingly we find that for all values of the input
parameter$\alpha$ the state does not violate the Bell's inequality
but when used as a teleportation channel can give a fidelity
higher than the classical optimum (i.e $\frac{2}{3}$).
\end{abstract}
\section{Introduction:}
The complementary theory of 'quantum no-cloning theorem' [1] is
the 'quantum no-deleting' principle [2]. It states that if we have
two identical qubits at the input port, then there does not exist
any linear map that will delete  unknown quantum state against a
copy. Quantum deletion [2] is more like reversible �uncopying� of
an unknown quantum state.  When memory in a quantum computer is
scarce, quantum deleting may play an important role. The
no-deleting principle does not prohibit us from constructing the
approximate deleting machine [3,4,5,6]. In [3] authors constructed
a input state state dependent deletion machine. Later in [5]
authors have constructed an universal deletion machine by making
different fidelities free from the probability amplitude of the
input state.\\
In his pioneering work , Bell [7] proved that, in general, two
quantum states cannot be considered as separate even if they are
located far from each other . When measurements are performed
independently on each of the systems , their results are
correlated in a way which cannot be explained by any local model.
Although Bell's inequalities witness entanglement but there are
entangled states which do not violate Bell's inequalities. Werner
[8] gave an example of an entangled state described by the density
operator $\rho_{W}=p|\psi^{-}\rangle\langle
\psi^{-}|+\frac{1-p}{4}I$, where
$|\psi^{-}\rangle=\frac{1}{\sqrt{2}}(|01\rangle-|10\rangle)$ and I
is the identity operator in the 4-dimensional Hilbert space, which
does not violate the Bell's inequality for
$\frac{1}{3}<p<\frac{1}{\sqrt2}$. Interestingly in this work, we
also find an example of an entangled state that does not violate
the Bell's inequality.\\
A milestone application of quantum information theory "Quantum
Teleportation" was proposed in [9]. The basic idea is to use a
pair of particles in a singlet state shared by distant partners
Alice and Bob to perform successful teleportation of an arbitrary
qubit from the sender Alice to the receiver Bob. Popescu [10]
noticed that the pairs in a mixed state could be still useful for
teleportation.  \\
A natural question arises in concern with teleportation whether
states which violate Bell-CHSH inequalities are suitable for
teleportation. Horodecki $\textit{et al}$ [11] showed that any
mixed two spin-$\frac{1}{2}$ state which
 violates the Bell-CHSH inequalities is suitable for
 teleportation. It was shown that for any state which violates Bell
 inequalities, $M(\rho)>1$, where $M(\rho)=max_{i>j}(u_{i}+u_{j})$ ,
 where $u_{i}$ are eigenvalues of the matrix
 $T^{\dag}T$ [11]. At this point Popescu raised an important question,
'What is the exact relationship between Bell's inequality
violation and teleportation? Since the exact relationship between
Bell's inequality and teleportation is unknown, it remains
interesting to see whether there exists any entangled state which
does not violate Bell's inequality but still can act as a
teleportation channel  with a fidelity $>\frac{2}{3}$ In our
recent work we have showed that the output entangled state of
Buzek-Hillery cloning machine [12] satisfy Bell's inequality and
at the same time can be used as a teleportation channel for
certain range of the machine parameter [13].\\
 Here in this work the basic motivation is to find such an entangled state which
 inspite of being consistent with the local realist model proposed by John Bell can
 exhibit the extreme non local phenomenon like teleportation .
 In that context we study the output of Pati-Braunstein deletion
machine [3] and show that this state can be a perfect example of
an entangled state that can act as a teleportation channel without violating Bell's inequality.
 In a nutshell here in this work we exemplify the existence of special class of
 entangled state which in spite of satisfying Bell's inequality can act as a
  teleportation channel. In other way round we analyze the output of Pati- Braunstein
  deletion machine and consequently establish its utility as a resource of
 quantum teleportation.
\section{ Pati-Braunstein Deletion Machine and output: }
Here we give a brief introduction to  Pati-Braunstein deletion
machine. Later we analyze its output and subsequently show that
this output
density can be utilized as a resource for faithful teleportation.\\
For orthogonal qubits, the action of such a machine is given by,
\begin{eqnarray}
&&|0\rangle|0\rangle|A\rangle\longrightarrow|0\rangle|\Sigma\rangle|A_0\rangle{}\nonumber\\&&
|0\rangle|1\rangle|A\rangle\longrightarrow|0\rangle|1\rangle|A\rangle{}\nonumber\\&&
|1\rangle|0\rangle|A\rangle\longrightarrow|1\rangle|0\rangle|A\rangle{}\nonumber\\&&
|1\rangle|1\rangle|A\rangle\longrightarrow|1\rangle|\Sigma\rangle|A_1\rangle
\end{eqnarray}
where $|A\rangle$ is the initial machine state and $|A_0\rangle$,
$|A_1\rangle$ are the final states of
ancilla.\\
$|\Sigma\rangle=m_1|0\rangle+m_2|1\rangle, (m_1^2+m_2^2=1)$
is the blank state.\\
For an arbitrary pair of qubits the action of such a
transformation is given by,
\begin{eqnarray}
|\Psi\rangle|\Psi\rangle|A\rangle=[\alpha^{2}|00\rangle+\beta^{2}|11\rangle+\alpha\beta(|0\rangle|1\rangle+
|1\rangle|0\rangle)]|A\rangle\nonumber\\\rightarrow\alpha^{2}|0\rangle|\Sigma\rangle|A_0\rangle+
\beta^{2}|1\rangle|\Sigma\rangle|A_1\rangle+\alpha\beta(|0\rangle|1\rangle+
|1\rangle|0\rangle)]|A\rangle\nonumber\\=|\Psi_{out}\rangle
\end{eqnarray}
Here the ancilla states $|A\rangle$ , $|A_0\rangle$  and
$|A_1\rangle$ are orthogonal to each other. The reduced density
matrix of the two qubits after the deletion operation is:
\begin{eqnarray}
\rho_{ab}=|\alpha|^4|0\rangle\langle
0|\otimes|\Sigma\rangle\langle \Sigma |+|\beta|^4|1\rangle\langle
1|\otimes|\Sigma\rangle\langle
\Sigma|+2|\alpha|^2|\beta|^2|\psi^{+}\rangle\langle \psi^{+}|
\end{eqnarray}
We now analyze the output (3) to investigate its inseparable
nature for different values of input parameters $\alpha$ and
$\beta$ and subsequently look for its efficiency as a
teleportation channel. For simplicity here we assume the initial
state $|\psi\rangle$ and blank state $|\Sigma\rangle$ to be a
member of real Hilbert Space, by taking $\alpha$,$\beta$,$m_{1}$
and $m_{2}$ as real quantities.\\
 \textbf{i) Inseparability of the output:} \\
The necessary and sufficient condition for the state $\rho$ of two
spin $\frac{1}{2}$ to be inseparable is that at least one of the
eigenvalues of the partially transposed operator defined as
$\rho^{T_B}_{m\mu,n\nu}=\rho_{m\nu,n\mu}$, is negative [14,15].
This is equivalent to the condition that at least one of the two
determinants. \\
$W_{3}= \begin{array}{|ccc|}
  \rho_{00,00} & \rho_{01,00} & \rho_{00,10} \\
  \rho_{00,01} & \rho_{01,01} & \rho_{00,11} \\
  \rho_{10,00} & \rho_{11,00} & \rho_{10,10}
\end{array}$ and $W_{4}=\begin{array}{|cccc|}
   \rho_{00,00} & \rho_{01,00} & \rho_{00,10} & \rho_{01,10}\\
  \rho_{00,01} & \rho_{01,01} & \rho_{00,11} & \rho_{01,11} \\
  \rho_{10,00} & \rho_{11,00} & \rho_{10,10} & \rho_{11,10} \\
  \rho_{10,01} & \rho_{11,01} & \rho_{10,11} & \rho_{11,11}
\end{array}$\\
is negative.\\
 After calculating the determinants for $\rho_{ab}$, we obtain the values of $W_3$
and $W_4$ as
\begin{eqnarray}
W_3=\alpha^6\beta^4m_1^2(\alpha^2+m_1^2\beta^2),~~~~W_4=-\alpha^6\beta^6[\alpha^4m_2^2+m_1^2\beta^4+\alpha^2\beta^2]
\end{eqnarray}
It is evident that $W_3>0~~~ (\forall \alpha,\beta,m_1,m_2$) and
$W_4<0~~~ (\forall \alpha,\beta,m_1,m_2$). Since at least one of
these two determinants is less than zero , hence we with full
generality conclude the state $\rho_{ab}$ to be inseparable.\\
\textbf{ii) Non-Violation of Bell's Inequality of two qubit
entangled
state:}\\
For simplicity we assume the blank state as
$|\Sigma\rangle=\frac{1}{\sqrt{2}}(|0\rangle+|1\rangle$), by
taking $m_1=m_2=\frac{1}{\sqrt{2}}$. It is a known fact that the
state which does not violate Bell's inequality must satisfy
$M(\rho)\leq1$, where $M(\rho)=\max_{i>j}(u_i+u_j)$ where $u_i$
and $u_j$ are the eigenvalues of $U=C^t(\rho)C(\rho)$
where $C(\rho)=[C_{ij}],C_{ij}=Tr[\rho\sigma_i\otimes \sigma_j]$[11]. \\
The eigenvalues of the matrix $U=C^t(\rho_{ab})C(\rho_{ab})$ for
the bipartite output state $\rho_{ab}$ of P-B deleting machine are
given by,
\begin{eqnarray}
&&u_1=4\alpha^4-8\alpha^6+4\alpha^8{}\nonumber\\&&
u_2=A+\frac{1}{2}\sqrt{B}{}\nonumber\\&&
u_3=A-\frac{1}{2}\sqrt{B}{}\nonumber\\&&  where
~~A=4\alpha^8-8\alpha^6+6\alpha^4-2\alpha^2+\frac{1}{2},{}\nonumber\\&&
B=1+64\alpha^{12}+224\alpha^8-8\alpha^2-192\alpha^{10}-128\alpha^6+40\alpha^4
\end{eqnarray}
A simple calculation will reveal that out of these three
eigenvalues, $u_1, u_2$ are the largest two. So taking these two
eigenvalues into consideration the expression for $M(\rho)$ will
be less than 1 for all values of $\alpha$ lying in the range
$(0,1)$.\\
\textbf{iii) Efficiency of the two qubit entangled state as
teleportation channel:}\\
Next we investigate whether the entangled state $\rho_{ab}$ can
act as a teleportation channel. Let us recall the eigenvalues of
the matrix $U=C^t(\rho_{ab})C(\rho_{ab})$ , $u_1$,$u_2$ and $u_3$
given in (5). Hence the teleportation fidelity $F_{max}$[11] is
given by,
\begin{eqnarray}
F_{max}=\frac{1}{2}[1+\frac{\sqrt{u_1}+\sqrt{u_2}+\sqrt{u_3}}{3}]
\end{eqnarray}
Next we find out the values of $F_{max}$ for different values of the input parameter $\alpha$
and enlist down in the following table.\\
{\bf TABLE 1:}\\
\begin{tabular}{|c|c|}
\hline  $\alpha$ & $F_{max}$  \\
\hline 0.1& 0.666783 \\
\hline 0.2 & 0.668531  \\
\hline 0.3 & 0.675915 \\
\hline 0.4 & 0.694094 \\
\hline 0.5 & 0.725347 \\
\hline 0.6 & 0.765805 \\
\hline 0.7 & 0.808094 \\
\hline 0.8 & 0.847683 \\
\hline 0.9 & 0.893974\\
\hline
\end{tabular}\\\\

Interestingly we find $F_{max} \geq   \frac{2}{3}  \forall \alpha
\in (0,1)$ .\\
{\bf A particular case : $\alpha=\beta=\frac{1}{\sqrt{2}}$}\\
Here we consider the case when the input state is an equal
superposition of qubits. For $\alpha=\beta=\frac{1}{\sqrt{2}}$,
the eigenvalues of U matrix are given by,
\begin{eqnarray}
u_1=u_2=u_3=\frac{1}{4}
\end{eqnarray}
The teleportation fidelity is
\begin{eqnarray}
F_{max}=\frac{3}{4}
\end{eqnarray}
Hence the output can be used as a teleportation channel $\forall
\alpha \in (0,1)$. Thus we see that the output of P-B deleting
machine can be used as a resource to quantum teleportation. Not
only that we see this output state gives a positive response to
our search for an entangled state that does not violate Bell's
inequality and still can act as a teleportation channel.
\section{Conclusion :}  In foundation of quantum mechanics violation of Bell's inequality
is considered to be a signature of the existence of non local
properties of the entangled states. At the same time quantum
teleportation (entangled states acting as teleportation channel
with a fidelity $>\frac{2}{3}$) itself is a manifestation of non
locality. Combining these two we can say that there is no such
elements of surprise in finding out entangled states that are
violating Bell's inequality and still can act as a teleportation
channel. But it remains interesting when we find out an entangled
state that does not violate Bell's inequality and still can act
as a teleportation channel. The basic motivation of the work is
to find out such an entangled state.  Here we have cited an
example of a two qubit entangled state (output of the
Pati-Braunstein deleting machine) [3] which does not violate
Bell-CHSH inequality [9] and at the same time can act as a useful
quantum channel for teleportation protocol. This work adds a
special feature to the Pati- Braunstein quantum deleting machine
by demonstrating its output state as a potential resource for
quantum teleportation.
\section{Acknowledgement:}
I.C acknowledges Prof C.G.Chakraborti for being the source of
inspiration in research work. N.G acknowledges his mother for her
love and blessings.
\section{Reference:}
$[1]$ W.K.Wootters and W.H.Zurek,Nature 299,802(1982).\\
$[2]$ A.K.Pati and S.L.Braunstein Nature 404, 164 (2000) .\\
$[3]$ A.K.Pati and S.L.Braunstein , arxiv:quant-ph/0007121v1(2000) \\
$[4]$ D.Qiu, Phys.Lett.A 301,112 (2002).\\
$[5]$ S.Adhikari, Phys. Rev. A 72, 052321 (2005).\\
$[6]$ S. Adhikari and B. S. Choudhury, Phys. Rev. A 73, 054303
(2006).\\
$[7]$ J.S.Bell , Physics 1 (1964) 195\\
$[8]$ R.F.Werner, Phys.Rev.A 40, 4277 (1989).\\
$[9]$ C.Benett,G.Brassard,C.Crepeau,R.Jozsa,A.peres and
W.K.Wootters, Phys Rev Lett 70(1993) 1895.\\
$[10]$ S.Popescu, Phys.Rev.Lett. 72, 797 (1994).\\
$[11]$ R.Horodecki, M.Horodecki, P.Horodecki, Phys.Lett.A 222, 21
(1996).\\
$[12]$ V. Buzek, M. Hillery, Phys. Rev.A 54, 1844 (1996).\\
$[13]$ S. Adhikari, N. Ganguly, I. Chakrabarty, B. S. Choudhury,
J. Phys. A: Math. Theor. 41 415302 (2008).\\
$[14]$ A.Peres, Phys.Rev.Lett. 77, 1413 (1996).\\
$[15]$ M.Horodecki, P.Horodecki, R.Horodecki, Phys.Lett.A 223, 1
(1996).\\
\end{document}